# 3D Inception-Based TransMorph: Pre- and Post-operative Multi-contrast MRI Registration in Brain Tumors

Javid Abderezaei[1*], Aymeric Pionteck[1*], Agamdeep Chopra[1], Mehmet Kurt[1]

[1] Department of Mechanical Engineering, University of Washington, Seattle, USA
* authors contributed equally to this paper

**Corresponding author:**
Mehmet Kurt, mkurt@uw.edu

**Abstract.** Deformable image registration is a key task in medical image analysis. The Brain Tumor Sequence Registration challenge (BraTS-Reg) aims at establishing correspondences between pre-operative and follow-up scans of the same patient diagnosed with an adult brain diffuse high-grade glioma and intends to address the challenging task of registering longitudinal data with major tissue appearance changes. In this work, we proposed a two-stage cascaded network based on the Inception and TransMorph models. The dataset for each patient was comprised of a native pre-contrast (T1), a contrast-enhanced T1-weighted (T1-CE), a T2-weighted (T2), and a Fluid Attenuated Inversion Recovery (FLAIR). The Inception model was used to fuse the 4 image modalities together and extract the most relevant information. Then, a variant of the TransMorph architecture was adapted to generate the displacement fields. The Loss function was composed of a standard image similarity measure, a diffusion regularizer, and an edge-map similarity measure added to overcome intensity dependence and reinforce correct boundary deformation. We observed that the addition of the Inception module substantially increased the performance of the network. Additionally, performing an initial affine registration before training the model showed improved accuracy in the landmark error measurements between pre and post-operative MRIs. We observed that our best model composed of the Inception and TransMorph architectures while using an initially affine registered dataset had the best performance with a median absolute error of 2.91 (initial error = 7.8). We achieved 6th place at the time of model submission in the final testing phase of the BraTS-Reg challenge.

## 1    Introduction

In many medical image analysis applications, deformable image registration is a key task. It is often used in a number of clinical applications such as image reconstruction [1], [2], segmentation [3], [4], motion tracking [5], [6], tumor growth monitoring, image guidance [7]–[9], and co-registration of MR brain images [10]. Using the deformation field, images of the same organ from different patients (inter-patient registration) or images of the same organ from the same person taken at different times or from different medical imaging equipment (intra-patient registration) are aligned for better visualization and comparison. The moving image is warped with the deformation field to match the fixed (target) image. Conventional image registration has been framed as an optimization problem to maximize a similarity measure that indicates the closeness of the moving image to the fixed image [11], [12]. Deep Convolutional Neural Networks (CNN) obtained significant success in several image analysis tasks such as object detection and recognition [13], [14], image classification [15], [16], and image segmentation [17], [18]. In image registration, several methods have been proposed to improve the performance in terms of registration accuracy and computation efficiency [19]–[24]. However, little attention has been given to the specific task of registering longitudinal brain Magnetic Resonance Imaging (MRI) scans containing pathologies. This task is particularly challenging due to major tissue appearance changes, and is still an unsolved problem.

Previous Brain Tumor Segmentation challenges (BraTS) were aimed at evaluating state-of-the-art methods for the segmentation of brain tumors. The Brain Tumor Sequence Registration (BraTS-Reg) challenge [25] aims at establishing correspondences between pre-operative and follow-up scans of the same patient diagnosed with an adult brain diffuse high-grade glioma. The BraTS-Reg challenge intends to establish a public benchmark environment for deformable registration algorithms. The associated dataset comprises de-identified multi-institutional multi-parametric MRI (mpMRI) data, curated for each scan's size and resolution, according to a common anatomical template. For each patient, a native precontrast (T1), a contrast-enhanced T1-weighted (T1-CE), a T2-weighted (T2) and a Fluid Attenuated Inversion Recovery (FLAIR) are provided. Clinical experts have generated extensive annotations of landmarks points within the scans descriptive of distinct anatomical locations across the temporal domain. A number of metrics such as Median Absolute Error (MAE), Robustness, and the Jacobian determinant are used to measure the performance of the algorithms proposed by participants.

In this work, inspired by the TransMorph [26] architecture, we propose a novel two-stage cascaded network. In the first stage, we use a variant of the Inception model [27] to merge the 4 image modalities together and extract the most relevant information from each type of the imaging contrasts. In the second stage, we use a variant of the TransMorph architecture to generate the displacement field used to warp the moving image.

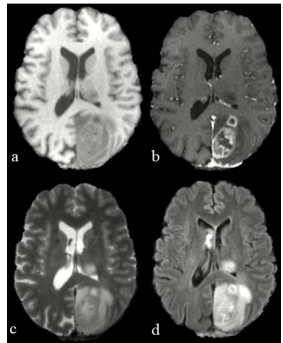

**Fig. 1.** Example of provided images, a) T1-weighted, b) T1-CE, c) T2-weighted, d) Flair

## 2 Methods

### 2.1 Model Architecture

An example image set of provided data is presented in Fig. 1, which shows a native precontrast (T1), a contrast-enhanced T1-weighted (T1-CE), a T2-weighted (T2) and a Fluid Attenuated Inversion Recovery (FLAIR) of one of the subjects. A cascaded network was developed by using 2 blocks derived from the Inception architecture [27] to combine the imaging modalities and the output was then fed into the TransMorph model [26] responsible for the image registration (Fig. 2).

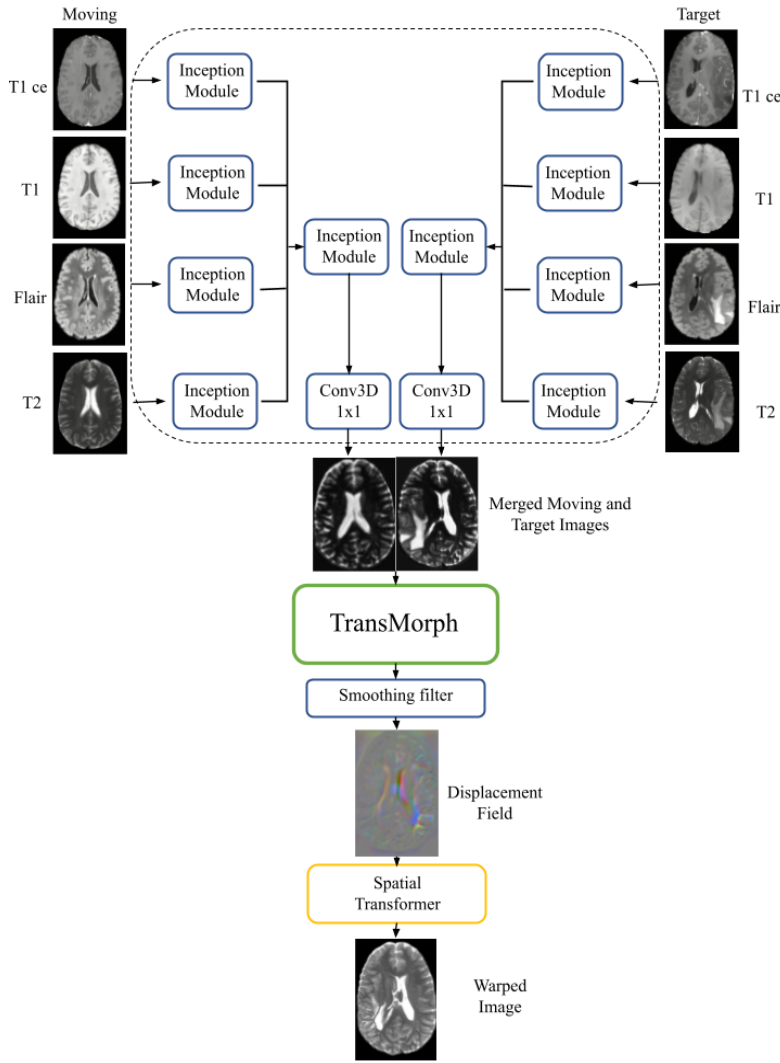

**Fig. 2.** Architecture of the proposed registration network

The Inception model [27] consists of 2 separate pipelines for both moving and target modalities (Fig. 2, 3). Each Inception pipeline consists of 4 input inception blocks corresponding to T1-CE, T1, Flair, and T2 as shown in Fig. 2. The outputs of the inception blocks are concatenated and passed through a sequential layer of another inception block followed by a single 3d convolution layer with kernel size = stride = 1, which outputs the desired single channel merged image. As described in [27], the main idea of the Inception architecture is finding out how an optimal local sparse structure in a convolutional vision network can be approximated and covered by readily available dense components. The goal is to find the optimal local construction and to repeat it spatially. We used the Inception module with dimension reductions described in Fig. 3 [27]. The advantage of the Inception module was that it allowed us to first process each contrast separately and extract the relevant information before concatenating them. The concatenated data was then passed through more Inception modules that merged the contrasts together and output new moving and target images. This approach had several advantages compared to a simple concatenation of the different contrast images. First, it added more training parameters corresponding to the data merging layers, which in the end should improve the results. It also helped reduce the memory requirements by merging 4 volumetric images into a single one.

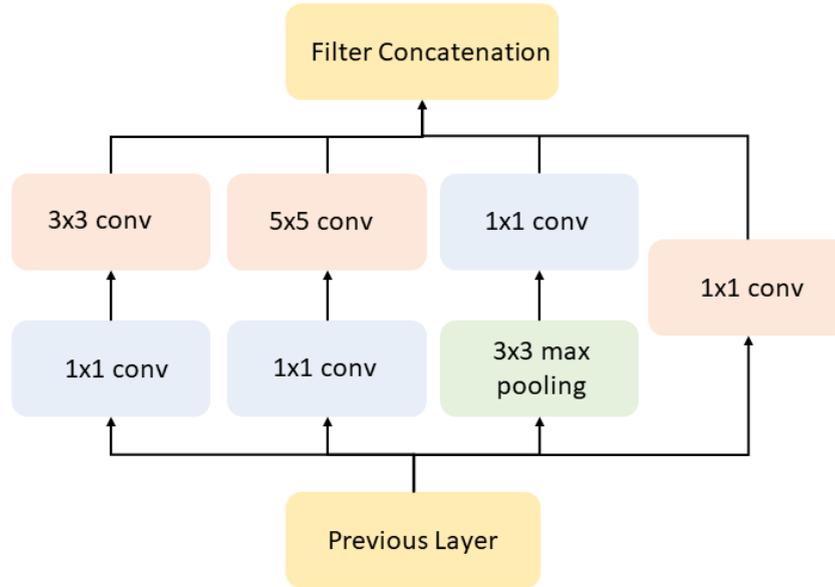

**Fig. 3.** Inception module with dimension reductions (adapted from [27])

The output of the Inception module was then passed to the TransMorph model as described in [26]. The output of the TransMorph model is a displacement field, which was then used as an input to a spatial transformer to register the moving images on the target images. TransMorph [26] is a hybrid TransformerConvNet framework designed for volumetric medical image registration. Transformer, initially designed for natural language processing tasks [28], has shown its potential in computer vision tasks. A Transformer determines which parts of the input sequence are essential based on contextual information by using self-attention mechanisms. Vision Transformer (ViT) [29], which applied the Transformer encoder directly to images, achieved state-of-the-art performance in image recognition. Transformer has also a strong potential for image registration as it can better comprehend the spatial correspondences between the moving and fixed images. Bridging of ViT and V-Net provided good performance in image registration [30]. In this method described in [26], the Swin Transformer [31] is employed as the encoder to capture the spatial correspondence between the input moving and fixed images. Then, a ConvNet decoder processed the information provided by the Transformer encoder into a dense displacement field. Long skip connections are used to maintain the flow of localization information between the encoder and decoder stages. Diffeomorphic variations of TransMorph were also introduced to ensure a smooth and topology-preserving deformation. The final output of the TransMorph model is a displacement field, which was used to warp the moving image (pre-operative) to match the target image (post-operative).

## 2.2 Training of the network

We performed multiple preprocessing steps on the input dataset before training the deep learning network. Because MRI intensity values are not normalized, we applied intensity normalization to each MRI modality of each patient independently by subtracting the mean and dividing by the standard deviation of the brain region. For each specific imaging modality, we also performed a histogram match between pre and post-surgery datasets. Finally, we divided our training into two trials. In the first trial, we trained the networks using the preprocessed dataset, while in the second trial, we first performed an affine registration between the pre and post-surgery images and then trained the networks.

Our network was implemented using PyTorch 1.1.1 [32]. The update of the weights of the network was done using Adam optimizer, with a batch size of 1 and a decaying learning rate with an initial value of $\alpha = 1e^{-4}$.

Training was performed on 2xNVIDIA A40 GPU 2x48GB memory. The initial loss function was based on a widely-used image similarity measure which computes the similarity between the deformed moving and the fixed images. We used the mean square error metric (Eq. 1), which was the mean of the squared differences in voxel (i) values between the moving (Y) and the target images ($\hat{Y}$).

$$L_{MSE} = \frac{1}{N}\sum_{i=1}^{N}(Y_i - \hat{Y}_i)^2 \qquad (1)$$

However, optimizing the similarity measure alone would encourage the warped images to be visually close to the target ones, but the displacement fields might not be realistic. Diffusion regularizer based on L2 loss [22] was added to impose smoothness in the displacement field. We also introduced a loss function based on edge detection to overcome intensity dependence and reinforce correct boundary deformation (Eq. (2-4), and Fig. 4). We first applied a 3x3x3 Gaussian kernel to blur the entire 3D input with σ = 1, and μ = 0. Then, we used 3 Sobel kernels (Eq. 2) [33], [34] to extract the edges as the magnitude of intensity gradients. Each Sobel kernel is a 3D filter that can be assimilated to a convolution filter with predefined weights. The weights of the 3 filters were chosen to detect edges in the x, y, and z axis, respectively (Eq. 2).

$$S_x = \left( \begin{pmatrix} -1 & 0 & 1 \\ -2 & 0 & 2 \\ -1 & 0 & 1 \end{pmatrix} \begin{pmatrix} -2 & 0 & 2 \\ -4 & 0 & 4 \\ -2 & 0 & 2 \end{pmatrix} \begin{pmatrix} -1 & 0 & 1 \\ -2 & 0 & 2 \\ -1 & 0 & 1 \end{pmatrix} \right)$$

$$S_y = \left( \begin{pmatrix} -1 & -2 & -1 \\ 0 & 0 & 0 \\ 1 & 2 & 1 \end{pmatrix} \begin{pmatrix} -2 & -4 & -2 \\ 0 & 0 & 0 \\ 2 & 4 & 2 \end{pmatrix} \begin{pmatrix} -1 & -2 & -1 \\ 0 & 0 & 0 \\ 1 & 2 & 1 \end{pmatrix} \right) \quad (2)$$

$$S_z = \left( \begin{pmatrix} -1 & -2 & -1 \\ -2 & -4 & -2 \\ -1 & -2 & -1 \end{pmatrix} \begin{pmatrix} 0 & 0 & 0 \\ 0 & 0 & 0 \\ 0 & 0 & 0 \end{pmatrix} \begin{pmatrix} 1 & 2 & 1 \\ 2 & 4 & 2 \\ 1 & 2 & 1 \end{pmatrix} \right)$$

These edges were located where there is a dramatic change in intensity within the filter. The final edge map was obtained by computing and normalizing the magnitude of the extracted intensity gradients (Eq. 3).

$$\varepsilon(A) = \sqrt{(S_x * A)^2 + (S_y * A)^2 + (S_z * A)^2} \quad (3)$$

The edge loss was calculated by comparing the edges of the predicted warped image to the target image edges using a standard mean square error loss function (Eq. 4).

$$L_{Edge} = \frac{1}{N} \sum_{i=1}^{N} (\varepsilon(Y_i) - \varepsilon(\hat{Y}_i))^2 \quad (4)$$

The outputs of each function were finally weighted and summed to calculate the final loss value (Eq. 5).

$$L_{Total} = L_{Edge} + L_{MSE} + L_{Diffusion} \quad (5)$$

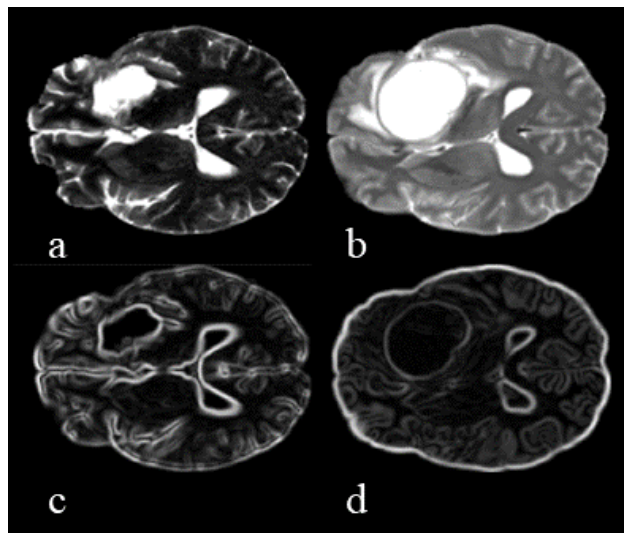

**Fig. 4.** Example of edge filter application, a) Initial moving image, b) initial target image, c) edges extracted from moving image, and d) edges extracted from target image

## 3      Results and Discussion

We analyzed 3 different variations of Inception-Transmorph for the task of image registration. We performed two sets of training on our dataset. In the first set of training, the raw dataset after normal preprocessing steps was used to train each model from scratch. In the second set of training, we initially performed an affine registration between pre and post-surgery datasets and then continued training similar to the first step. Finally, we selected the best model according to its performance on the evaluation data. The evaluation of the results was performed on an online evaluation platform (https://ipp.cbica.upenn.edu/). Each model was trained on 140 multi-modal MRI datasets composed of T1, T1-CE, T2, and Flair images. After training the models, we observed that it can successfully warp the moving images into the target ones (Fig. 5). Transmorph + Inception after initial affine registration of the dataset was our best-performing model, which was able to decrease the initial landmark median absolute error of 7.8 mm to 2.91 mm (Table. 1). When the dataset was not initially affine registered, we found that the combination of Inception and TransMorph with the addition of edge maps in the training had the best performance compared to the other models (Table. 1). Adding the edge maps in the loss function improved the results compared to initial TransMorph and TransMorph + Inception, by 15.6% and 21.8% respectively. The addition of the edge maps also dramatically improved the robustness results (about 25% increased performance; Table. 1). This is primarily due to the edge map's ability to enforce the boundary of the deformation and hence more accurately drive the displacement direction.

In the second set of training where we initially performed an affine registration and then trained the models, we observed that the models perform substantially better than without affine registration. Here, TransMorph + Inception had the best performance with a landmark median absolute error of 2.91 mm which is about 28% better than the best model without the initial affine registration (Table. 1). This model also had the highest robustness of 0.82 (Table. 1). One interesting observation we had was that in this set of training as opposed to the previous one (after affine registration), the use of edge maps in the loss function decreased the performance of the model (Table. 1). One possible explanation is that the edge maps contribute more to improving the accuracy of affine registration rather than the general nonlinear warping of the images.

**Table 1.** Evaluation of the trained models with the BraTS-Reg 2022 datasets

|  |  | Median Absolute Error (mm) | Mean Absolute Error (mm) | Robustness |
|---|---|---|---|---|
| Without Affine Registration | TransMorph | 7.29 | 7.82 | 0.57 |
|  | TransMorph + Inception | 7.66 | 8.09 | 0.59 |
|  | TransMorph + Inception + Edge Maps | 6.15 | 6.48 | 0.76 |
| **With Affine Registration** | TransMorph | 4.06 | 5.09 | 0.72 |
|  | **TransMorph + Inception** | **2.91** | **3.62** | **0.82** |
|  | TransMorph + Inception + Edge Maps | 3.51 | 4.00 | 0.79 |

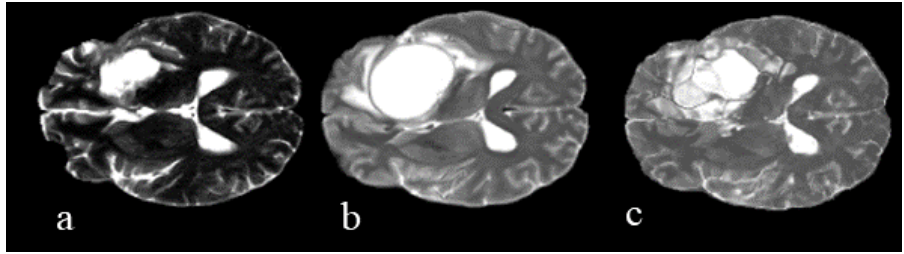

**Fig. 5.** Example of warped image with output displacement field of the network: a) Moving image, b) target image, c) warped image

## 4     Conclusion

In this paper, we proposed a two-stage cascaded network. Our approach first merged the 4 image modalities through an Inception model to extract the most relevant features from the different contrasts. Then the output of the first stage went through a model adapted from TransMorph to extract the displacement field. Experiments on the BraTS-Reg 2022 validation set demonstrated that our method could obtain results very close to the target. Adding the edge maps in the loss function improved the results only when we trained the model without using the affine-registered dataset. However, the inclusion of the edge maps resulted in worse performance when we used an affine registration prior to training the model. This is probably due to the effectiveness of the edge maps loss function in helping with the affine registration rather than the nonlinear deformation of the images. We believe that with further fine-tuning of hyperparameters as well as a higher number of training epochs, and adopting a multi-scale approach to account for localized deformations, we can achieve improved results with the introduced Inception-Transmorph network.

[9] T. De Silva et al., "3D–2D image registration for target localization in spine surgery: investigation of similarity metrics providing robustness to content mismatch," Phys. Med. Biol., vol. 61, no. 8, p. 3009, 2016.

[10] C. Gaser, "Structural MRI: Morphometry," in Neuroeconomics, M. Reuter and C. Montag, Eds. Berlin, Heidelberg: Springer, 2016, pp. 399–409. doi: 10.1007/978-3-642-35923-1_21.

[11] J. Ashburner, "A fast diffeomorphic image registration algorithm," NeuroImage, vol. 38, no. 1, pp. 95–113, Oct. 2007, doi: 10.1016/j.neuroimage.2007.07.007.

[12] B. B. Avants, C. L. Epstein, M. Grossman, and J. C. Gee, "Symmetric diffeomorphic image registration with cross-correlation: Evaluating automated labeling of elderly and neurodegenerative brain," Med. Image Anal., vol. 12, no. 1, pp. 26–41, Feb. 2008, doi: 10.1016/j.media.2007.06.004.

[13] J. Li, X. Liang, S. Shen, T. Xu, J. Feng, and S. Yan, "Scale-Aware Fast R-CNN for Pedestrian Detection," IEEE Trans. Multimed., vol. 20, no. 4, pp. 985–996, Apr. 2018, doi: 10.1109/TMM.2017.2759508.

[14] Y. Wu et al., "Rethinking Classification and Localization for Object Detection," 2020, pp. 10186–10195. Accessed: Jul. 15, 2022. [Online]. Available: https://openaccess.thecvf.com/content_CVPR_2020/html/Wu_Rethinking_Classification_and_Localization_for_Object_Detection_CVPR_2020_paper.html

[15] A. Krizhevsky, I. Sutskever, and G. E. Hinton, "ImageNet Classification with Deep Convolutional Neural Networks," in Advances in Neural Information Processing Systems, 2012, vol. 25. Accessed: Jul. 15, 2022. [Online]. Available: https://proceedings.neurips.cc/paper/2012/hash/c399862d3b9d6b76c8436e924a68c45b-Abstract.html

[16] Q. Xie, M.-T. Luong, E. Hovy, and Q. V. Le, "Self-Training With Noisy Student Improves ImageNet Classification," in 2020 IEEE/CVF Conference on Computer Vision and Pattern Recognition (CVPR), Seattle, WA, USA, Jun. 2020, pp. 10684–10695. doi: 10.1109/CVPR42600.2020.01070.

[17] A. Tao, K. Sapra, and B. Catanzaro, "Hierarchical multi-scale attention for semantic segmentation," ArXiv Prepr. ArXiv200510821, 2020.

[18] L.-C. Chen, Y. Yang, J. Wang, W. Xu, and A. L. Yuille, "Attention to Scale: Scale-Aware Semantic Image Segmentation," 2016, pp. 3640–3649. Accessed: Jul. 15, 2022. [Online]. Available: https://openaccess.thecvf.com/content_cvpr_2016/html/Chen_Attention_to_Scale_CVPR_2016_paper.html

[19] X. Yang, R. Kwitt, M. Styner, and M. Niethammer, "Quicksilver: Fast predictive image registration – A deep learning approach," NeuroImage, vol. 158, pp. 378–396, Sep. 2017, doi: 10.1016/j.neuroimage.2017.07.008.

[20] A. V. Dalca, G. Balakrishnan, J. Guttag, and M. R. Sabuncu, "Unsupervised Learning for Fast Probabilistic Diffeomorphic Registration," arXiv, 2018, Accessed: Jul. 15, 2022. [Online]. Available: https://dspace.mit.edu/handle/1721.1/137585

[21] G. Balakrishnan, A. Zhao, M. R. Sabuncu, J. Guttag, and A. V. Dalca, "An Unsupervised Learning Model for Deformable Medical Image Registration," ArXiv180202604 Cs, Apr. 2018, doi: 10.1109/CVPR.2018.00964.

[22] G. Balakrishnan, A. Zhao, M. R. Sabuncu, J. Guttag, and A. V. Dalca, "VoxelMorph: A Learning Framework for Deformable Medical Image Registration,"